\documentclass[11pt]{article}
\usepackage{epsfig}
\usepackage{mathtext}      
\usepackage{amsmath,amssymb} 

\newcommand{\be}{\begin{equation}}
\newcommand{\ee}{\end{equation}}
\newcommand{\bea}{\begin{eqnarray}}
\newcommand{\eea}{\end{eqnarray}}
\newcommand{\ba}{\begin{eqnarray*}}
\newcommand{\ea}{\end{eqnarray*}}

\def\({\left(}
\def\){\right)}

\def\f21{{_2F_1}}

\def\al{\alpha}
\def\ap{\alpha'}
\def\bp{\beta'}
\def\bt{\beta}

\def \to   {\mbox{$\rightarrow$}}

\def\ga{\gamma}
\def\gp{\gamma'}
\def\de{\delta}

\newcommand{\Fe}[2]{\Fs{#1}{#2}{z}}
\newcommand{\Fey}[2]{\Fs{#1}{#2}{y}}
\newcommand{\FxC}[2]{\Fs{#1}{#2}{\frac{y}{1-x} }}
\newcommand{\FxP}[2]{\Fs{#1}{#2}{\frac{x}{1-y} }}
\newcommand{\FXY}[2]{\Fs{#1}{#2}{\frac{1-x}{1-y} }}
\newcommand{\Fxy}[2]{\Fs{#1}{#2}{\frac{x}{y} }}

\newcommand{\Fh}[2]{\,{}_#1F_#2}
\newcommand{\Fs}[3]{\!\!\left[\begin{array}{c}#1\,;\\#2\,;\end{array}#3\right]}

\newcommand{\Ftt}[2]{\Fs{#1}{#2}{\theta  }}
\newcommand{\FTT}[2]{\Fs{#1}{#2}{\frac{4m^2u}{(s-4m^2)(t-4m^2)}}}

\newcommand{\FXV}[2]{\Fs{#1}{#2}{x v}}

\newcommand{\Fis}[2]{\Fs{#1}{#2}{\frac{s+t}{s}}}
\newcommand{\Fit}[2]{\Fs{#1}{#2}{\frac{s+t}{t}}}

\newcommand{\Fuy}[2]{\Fs{#1}{#2}{ u y}}

\newcommand{\Fus}[2]{\Fs{#1}{#2}{\frac{4m^2u}{(t+u)(s+u)}}}

\newcommand{\FHQ}[2]{\Fs{#1}{#2}{ 1+\frac{(s-m^2)^2}{st} }}
\newcommand{\Fsg}[2]{\Fs{#1}{#2}{ \sigma }}
\newcommand{\FSX}[2]{\Fs{#1}{#2}{ 1+\frac{(x-m^2)^2}{xt} }}
\newcommand{\FSW}[2]{\Fs{#1}{#2}{ v }}

\newcommand{\FHH}[2]{\Fs{#1}{#2}{ 1+\frac{st}{(t-m^2)^2} }}
\newcommand{\FBB}[2]{\Fs{#1}{#2}{ 1+\frac{s-4m^2}{t} }}
\newcommand{\FmW}[2]{\Fs{#1}{#2}{ -\frac{ts}{w} }}

\begin{document}

\begin{titlepage}
\thispagestyle{empty}
\onecolumn

\begin{flushleft}
DESY~09--054  \\
\end{flushleft}

\vspace*{0.2cm}
\begin{center}
{\bf \Large
  Functional equations for one-loop master integrals
\\
for heavy-quark production and Bhabha scattering}
\\

\vspace{2cm}

{Bernd A. Kniehl and Oleg V. Tarasov\footnote{On leave
of absence from Joint Institute for Nuclear Research,
141980 Dubna (Moscow Region) Russia.}        \\
{\normalsize\it II. Institut f\"ur Theoretische Physik, 
Universit\"at Hamburg,}\\
{\normalsize\it Luruper Chaussee 149, 22761 Hamburg, Germany}}

\end{center}

\vspace*{1.0cm}

\begin{abstract}
The method for obtaining functional equations, recently proposed 
by one of the authors \cite{Tarasov:2008hw}, is applied to one-loop 
box integrals needed in calculations of radiative corrections to 
heavy-quark production and Bhabha scattering. We present relationships 
between these integrals with different arguments and box integrals 
with all propagators being massless. It turns out that functional 
equations are rather useful for finding imaginary parts and performing 
analytic continuations of Feynman integrals.  For  the box master integral needed 
in Bhabha scattering, a new representation in  terms of hypergeometric functions
admitting one-fold integral representation  is derived. 
The hypergeometric representation of a master integral for 
heavy-quark production follows from the functional equation.

\medskip


\end{abstract}

\end{titlepage}

\newpage

\section{Introduction}

The production of heavy quarks at hadron colliders has become
a very important field of research. The very large production 
rates for both top and bottom  quarks at the CERN Large Hadron 
Collider (LHC) will allow for studies  of  heavy quarks with  high 
precision. The full next-to-leading-order (NLO) radiative  
corrections to the hadroproduction of  heavy flavors was completed 
in Ref.~\cite{Nason:1987xz}. The theoretical NLO 
predictions suffer  from the usual uncertainty  resulting from 
the freedom in the  choice of renormalization and factorization 
scales of perturbative QCD. To reduce such  uncertainties, 
next-to-next-to-leading order  (NNLO) calculations are needed. 
The computation of NNLO corrections is complicated due to  great 
technical difficulties, mainly related to  the evaluation of two-loop
Feynman integrals. To overcome such difficulties, it is of great 
importance to develop new methods and approaches for calculating 
Feynman integrals. The main complications of these calculations 
are related to  the fact that the integrals depend on  several 
kinematical variables. As was noted in Ref.~\cite{Tarasov:2006nk},
the most appropriate methods for calculating such integrals may 
be those based on a different kind of recurrence relations. Such
methods  can be based on the solution of recurrence relations with 
respect to the exponent of a propagator in the  integral 
\cite{Kazakov:1983pk} or on the solution of dimensional recurrences 
\cite{Tarasov:1996br,Tarasov:2000sf}.

A significant simplification of the computation of Feynman 
integrals depending on  several kinematical variables may be
achieved by using a new type of relationships between Feynman
integrals through functional equations with respect to kinematical
variables as proposed in Ref.~\cite{Tarasov:2008hw}. As was shown in 
Ref.~\cite{Tarasov:2008hw},  Feynman integrals with several 
kinematical variables can be expressed in terms of integrals 
with a lesser number of variables, which   significantly
simplifies their evaluation.

It is the purpose of the present paper to apply the general method 
for finding functional equations  \cite{Tarasov:2008hw} to 
integrals required in calculations of radiative corrections to
important physical processes and to use those relations for the analytic 
computation of  these  integrals.

Our paper is organized as follows. 
In Section 2, 
we give definitions and notations. 
In Section 3, 
we present functional equations for the on-shell master integrals
from heavy-quark production and Bhabha scattering. 
In Section 4, 
a new hypergeometric representation in terms of the Appell functions 
$F_1$ and $F_3$ and the Gauss hypergeometric function $_2F_1$ for
the one-loop box integral from Bhabha scattering is presented.
In Section 5, 
we describe how to use functional equations to find imaginary 
parts of the considered integrals. New analytic results for the 
imaginary parts are presented. Using dispersion relations, we write 
also one-fold integral representations for real parts of integrals. 
In Section 6,
we present functional equations for master integrals
from heavy-quark production with one quark leg off shell.

\section{Definitions and notations}

As was shown in Ref.~\cite{Tarasov:2008hw}, functional equations
for  one-loop  integrals corresponding to diagrams with four 
external legs can be derived,  for example,  from the following 
equation  obtained in  Refs.~\cite{Tarasov:1996br,Fleischer:1999hq}:
\begin{eqnarray}
\lefteqn{G_4~{\bf j^+}
 I_5^{(d+2)}(m_1^2,m_2^2,m_3^2,m_4^2,m_5^2;\{s_{kr}\} )
    - (\partial_j \Delta_5)
 I_5^{(d)}(m_1^2,m_2^2,m_3^2,m_4^2,m_5^2;\{s_{kr}\})} 
\nonumber
\\
&=&
(\partial_j \partial_1 \Delta_5)
~ I_4^{(d)}(m_2^2,m_3^2,m_4^2,m_5^2; 
s_{23},s_{34},s_{45},s_{25};~s_{35},s_{24})
\nonumber 
\\
&&{}+
(\partial_j \partial_2 \Delta_5)
~I_4^{(d)}(m_1^2,m_3^2,m_4^2,m_5^2;
s_{13},s_{34},s_{45},s_{15};~s_{35},s_{14})
\nonumber
\\
&&{}+
(\partial_j\partial_3 \Delta_5)
~I_4^{(d)}(m_1^2,m_2^2,m_4^2,m_5^2 ;
s_{12},s_{24},s_{45},s_{15};~s_{25},s_{14})
\nonumber
\\
&&{}+
(\partial_j\partial_4 \Delta_5)
~I_4^{(d)}(m_1^2,m_2^2,m_3^2,m_5^2 ;
 s_{12},s_{23},s_{35},s_{15};~s_{25},s_{13})
\nonumber
\\
&&{}+
(\partial_j\partial_5 \Delta_5)
~I_4^{(d)}(m_1^2,m_2^2,m_3^2,m_4^2 ;
s_{12},s_{23},s_{34},s_{14};~s_{24},s_{13}),
\label{I5intoI4}
\end{eqnarray}
where the operator   ${\bf j^{+}}$  shifts the indices 
$\nu_j \to \nu_{j } + 1$, $G_4$ is the Gram determinant, 
and $\Delta_5 $ is the modified Cayley determinant, defined as
\begin{equation}
G_{4}= -16 \left|
\begin{array}{cccc}
  p_{15}p_{15}   & p_{15}p_{25} & p_{15}p_{35} & p_{15}p_{45} \\
  p_{15}p_{25}   & p_{25}p_{25} & p_{25}p_{35} & p_{25}p_{45} \\
  p_{15}p_{35}   & p_{25}p_{35} & p_{35}p_{35} & p_{35}p_{45} \\
  p_{15}p_{45}   & p_{25}p_{45} & p_{35}p_{45} & p_{45}p_{45} 
\end{array}
\right|,
\qquad
\Delta_5=  \left|
\begin{array}{ccccc}
Y_{11} & Y_{12}  & Y_{13} & Y_{14}  & Y_{15} \\
Y_{12} & Y_{22}  & Y_{23} & Y_{24}  & Y_{25} \\
Y_{13} & Y_{23}  & Y_{33} & Y_{34}  & Y_{35} \\
Y_{14} & Y_{24}  & Y_{34} & Y_{44}  & Y_{45} \\
Y_{15} & Y_{25}  & Y_{35} & Y_{45}  & Y_{55} 
\end{array}
         \right|,
\end{equation}
\begin{equation}
Y_{ij}=m_i^2+m_j^2-s_{ij}, 
\qquad  s_{ij}=p_{ij}^2, 
\qquad  p_{ij}=p_i-p_j,
\qquad  \partial_j = \frac{\partial }{ \partial m_j^2}.
\end{equation}
The integral  $I_5^{(d)}(\{m_j^2\},\{s_{kr}\}) $ corresponds to a
diagram with five external legs and the integrals
$I_4^{(d)}(\{m_j^2\},\{s_{kr}\}) $ in Eq.~(\ref{I5intoI4}) are 
defined as 
\begin{eqnarray}
\lefteqn{I_4^{(d)}(m_n^2,m_j^2,m_k^2,m_l^2;~
s_{nj},s_{jk},s_{kl},s_{nl};~s_{jl},s_{nk})}
\nonumber
\\
&=&
\int \frac{d^dq}{i\pi^{d/2}}~\frac{1}{[(q-p_n)^2-m_n^2]
[(q-p_j)^2-m_j^2][(q-p_k)^2-m_k^2][(q-p_l)^2-m_l^2]}.
\end{eqnarray}
The diagram corresponding to the integral 
$I_4^{(d)}(m_1^2,m_2^2,m_3^2,m_4^2;~\{s_{ij}\})$ is presented in Fig.~1.
\begin{figure}[h]
\begin{center}
\includegraphics[scale=0.9]{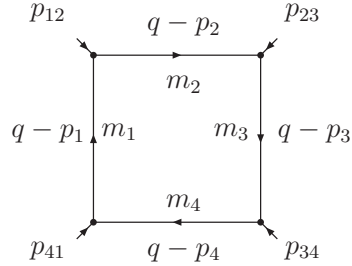}
\end{center}
\caption{\it  Diagram corresponding to the integral  
$I_4^{(d)}(m_1^1,m_2^2,m_3^2,m_4^2;~
s_{12},s_{23},s_{34},s_{14};~s_{24},s_{13}).$}
\end{figure}

In what follows,  we will use the following short-hand notation 
for integrals needed in the calculation of the one-loop radiative 
corrections to the process $e^+e^- \rightarrow e^+e^-$, the 
so-called Bhabha scattering \cite{Bhabha:1936zz}, and heavy-quark production:
\begin{eqnarray}
B(s,t)&=&I_4^{(d)}(0,m^2,0,m^2;~m^2,m^2,m^2,m^2;~s,t),
\nonumber \\
D_2(s,t)&=&I_4^{(d)}(0,0,0,m^2;~0,0,m^2,m^2;~t,s).
\end{eqnarray}
The diagrams corresponding to these integrals are depicted in Fig.~2.
\begin{figure}[h]
\begin{center}
\includegraphics[width=13cm]{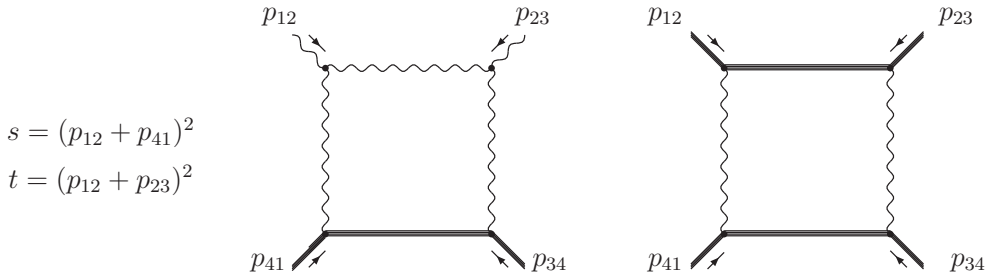}

\end{center}
\caption{\it  Diagrams corresponding to the integrals  $D_2(t,s)$ 
and $B(s,t)$. Lines with a zero internal mass $m_i=0$ (for
internal lines) or a zero virtuality $s_{ij}=0$ (for external
lines) are shown wavy. Solid lines have a non-zero internal mass 
or a non-zero virtuality.}
\end{figure}

\section{Functional equations for the integrals
 \boldmath{$B(s,t)$} and \boldmath{$D_2(t,s)$} }

In this section, we  present in detail the derivation of
functional equations for the scalar integrals $B(s,t)$ and 
$D_2(t,s)$. As was proposed in Ref.~\cite{Tarasov:2008hw}, one can 
obtain a functional equation for the integral 
$I_4^{(d)}(\{m_j^2\},\{s_{kr}\}) $  by eliminating terms with 
$I_5^{(d)}(\{m_j^2\},\{s_{kr}\}) $ from Eq.~(\ref{I5intoI4}) through an 
appropriate choice of kinematical variables.  The integral 
$I_5^{(d)}(\{m_j^2\},\{s_{kr}\}) $ depends on 15 kinematical variables, 
while the integral   $I_4^{(d)}(\{m_j^2\},\{s_{kr}\}) $ depends on 10 
variables. Therefore, to obtain a functional equation for the integral 
$I_4^{(d)}(\{m_j^2\},\{s_{kr}\}) $ with all 10  kinematical
variables arbitrary, we can impose conditions on some 5 variables.
To eliminate terms with $I_5^{(d)}(\{m_j^2\},\{s_{kr}\})$ from
Eq.~(\ref{I5intoI4}), two equations must be fulfilled:
\begin{equation}
G_4=0, \qquad\partial_j\Delta_5=0,
\label{MUST_BE_SO}
\end{equation}
thus fixing two kinematical variables. There are several options to use 
the remaining three  variables. One option is to set these variables 
to some particular values and obtain a functional equation connecting 
the integral of interest  with integrals which are easy to evaluate 
and/or integrals similar to the original one but with different 
kinematics. 
Another option to choose variables is to reduce the number of terms 
in the functional equation by requiring some derivatives
$\partial_i \partial_j \Delta_5$ to be zero.
Also,  one can use some combination of these two options.
The derivation of the numerous functional equations corresponding
to the different options and their combinations described above
was done on a computer. We describe the most useful functional
equations below.

To be definite, let us assume that our integral of interest is the 
last integral on the right-hand side of Eq.~(\ref{I5intoI4}). 
If we set in Eq.~(\ref{I5intoI4})  $m_1=m_3=0$, $m_2^2=m_4^2=m^2$, 
$s_{12}=s_{23}=s_{34}= s_{14} =m^2$, and $s_{24}=s$, $s_{13}=t$, then the 
last integral on the right-hand side of this equation   corresponds 
to our integral $B(s,t)$.

Setting $m_5^2=0$ and choosing different particular values of the 
remaining  two kinematical  variables and/or   requiring some second 
derivatives  $\partial_i \partial_j \Delta_5$ to be zero, one 
can get from Eq.~(\ref{I5intoI4}) rather different functional equations
for the integral $B(s,t)$. For some specific choice of kinematical 
variables, also the integral $D_2(s,t)$  appears in the functional 
equation. By imposing different conditions, we try to find 
equations connecting the integrals $B(s,t)$ and $D_2(s,t)$ with simpler
integrals, for example, with integrals having more massless propagators 
and  simpler external kinematics. 
We would like to note that, in the present investigation, we
always set $m_5^2=0$ in order to avoid the appearance of integrals
with three propagators having nonzero mass in the functional
equations.

Substituting $j=2$, $m_2^2=m_3^2=m_5^2=0$, $s_{12}=s_{23}=s_{34}=s_{14}=m^2$,
$s_{24}=s$, and $s_{13}=t$ into Eq.~(\ref{I5intoI4}) 
and choosing  $s_{15}$, $s_{25}$, $s_{35}$, and $s_{45}$  from the conditions
\begin{equation}
G_4=0,\qquad\partial_2 \Delta_5 =0,
\qquad\partial_1 \partial_2 \Delta_5 =0,
\qquad\partial_2 \partial_3 \Delta_5 =0,
\end{equation}
we arrive at the following equation:
\begin{equation}
B(s,t) = \frac{m^2}{s}(1+\alpha_+)D_2(t,m^2\alpha_{+})
+\frac{m^2}{s}(1+\alpha_{-}) D_2(t,m^2\alpha_{-}),
\label{bb_into_gluon}
\end{equation}
where
\begin{equation}
\alpha_{\pm}= \frac{1 \pm \beta_{s}}{1 \mp \beta_{s}},
\qquad\beta_s = \sqrt{1-\frac{4m^2}{s}}.
\end{equation}
Thus,  we have a relation connecting  the  integral $B(s,t)$
with an integral having only one massive propagator, i.e.\ with the
integral $D_2(s,t)$. It turns out that  the integral  $D_2(s,t)$ 
in Eq.~(\ref{bb_into_gluon}) satisfies the  following functional  
equation:
\begin{equation}
D_2(t,s)=\frac{m^2}{s} ~D_2\left(t,\frac{m^4}{s}\right)
+\frac{s-m^2}{s}~
I_4^{(d)}\left(0,0,0,0;~0,0,0,0;~\frac{(s-m^2)^2}{s},t\right),
\label{analitic_cont}
\end{equation}
which can be obtained from  Eq.~(\ref{I5intoI4}) by setting $j=5$,
$m_1^2=m_2^2=m_3^2=m_5^2=s_{12}=s_{23}=0$, 
$s_{34}=s_{14}=m_4^2=m^2$, and $s_{24}=s,~s_{13}=t$
and imposing the conditions
\begin{equation}
G_4 = \partial_5\Delta_5=
\partial_1 \partial_5\Delta_5=\partial_3 \partial_5\Delta_5=0.
\end{equation}
The last integral in Eq.~(\ref{analitic_cont}) corresponds to the box 
integral with all propagators massless and the squares of all external
momenta equal to zero. By using Eq.~(\ref{analitic_cont}) 
and taking into account the relation $\alpha_{+}\alpha_{-}=1$, 
one can write the integral in Eq.~(\ref{bb_into_gluon}) 
with argument  $\alpha_{+}$ as
\begin{equation}
D_2(t,m^2 \alpha_{+})= \alpha_{-}~D_2(t,m^2 \alpha_{-})
+(1-\alpha_{-})~
I_4^{(d)}\left(0,0,0,0;~0,0,0,0;~s-4m^2,t\right).
\label{d2_ac}
\end{equation}
Substituting this relation into Eq.~(\ref{bb_into_gluon}), gives
\begin{equation}
B(s,t) =(1-\beta_s)D_2(t,m^2 \alpha_{-})
+\beta_s
~I_4^{(d)}(0,0,0,0;~0,0,0,0;~s-4m^2,t).
\label{alpha_minus}
\end{equation}
We illustrate this relation in Fig.~3.
From Eq.~(\ref{alpha_minus}), the master integral from heavy-quark
production is found to be
\begin{equation}
D_2(s,t)=
\frac{t+m^2}{2t}~B\left(\frac{(t+m^2)^2}{t},s\right)
+\frac{t-m^2}{2t}~
I_4^{(d)}\left(0,0,0,0;~0,0,0,0;~\frac{(t-m^2)^2}{t},s\right).
\label{D2_B_I4}
\end{equation}
Here, we would like to remark that,  for $\varepsilon=(4-d)/2 \rightarrow 0$, 
the integral $D_2$ has a pole proportional to $1/\varepsilon^2$, while the
leading singularity of the integral $B(s,t)$ is $1/\varepsilon$. The leading
$1/\varepsilon^2$ singularity on the right-hand side comes from
the massless integral $I_4$.
\begin{figure}[h]
\begin{center}
\includegraphics[scale=1.2]{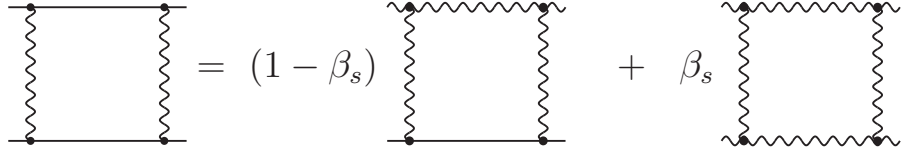}
\end{center}
\caption{A schematic depiction of Eq.~(\ref{alpha_minus}).
Wavy lines correspond 
to massless scalar propagators and solid  lines to massive
propagators.}
\end{figure}

Analytic formulae for both integrals on the right-hand side of 
Eq.~(\ref{D2_B_I4}) are given in the next sections.

\section{Analytic result for the integral \boldmath{$B(s,t)$}}

Using the  method of dimensional recurrences 
\cite{Tarasov:1996br,Tarasov:2000sf}, the  following hypergeometric 
representation for the integral $B(s,t)$ was obtained in 
Ref.~\cite{Fleischer:2003rm}:
\begin{eqnarray}
B(s,t)&=& \frac{(-2)}{mt(s-4m^2)}I_2^{(d)}(0,0;~t)~
F_1\left(\frac{d-3}{2},1,\frac12; \frac{d-1}{2};\frac{tz}{4},
-\frac{t \theta}{4m^2} \right)
\nonumber \\
&&{}+\frac{2(2-d)}{t(s-4m^2)}I_2^{(d)}(0,m^2;~0) ~
\nonumber \\
&&{}\times \left[
F_2\left(\frac{d-3}{2},1,1,\frac32, \frac{d-2}{2};
 \frac{s}{s-4m^2}, -m^2z \right)
-\frac{1}{d-3}\phi\left(-m^2z,\theta \right)
\right],
\label{solution}
\end{eqnarray}
where
\begin{equation}
z=\frac{4u}{t(4 m^2-s)},\qquad\theta=1-\frac{4m^2}{t},\qquad u=4m^2-s-t,
\label{z_teta}
\end{equation}
and $I_2^{(d)}$ are the one-loop propagator type integrals
\begin{eqnarray}
I_2^{(d)}(0,m^2;~0)&=&
\frac{1}{i \pi^{d/2}} \int \frac{d^dk_1}{k_1^2(k_1^2-m^2)}=
- \Gamma\left(1-\frac{d}{2}\right) m^{d-4},
\nonumber\\
I_2^{(d)}(0,0;~p^2)&=&
\frac{1}{i \pi^{d/2}} \int \frac{d^dk_1}{k_1^2(k_1-p)^2}~=~
 \frac{-\pi^{\frac32}~(-p^2)^{\frac{d}{2}-2}}
{2^{d-3}  \Gamma\left(\frac{d-1}{2}\right)
 \sin \frac{\pi d }{2}}.
\label{I2_00p}
\end{eqnarray}
Here, the Appell hypergeometric functions are
\begin{eqnarray}
F_1\left(\frac{d-3}{2},1,\frac{1}{2},\frac{d-1}{2};x,y\right)&=&
\sum_{r=0}^{\infty} \sum_{s=0}^{\infty} 
\frac{\left( \frac{d-3}{2} \right)_{r+s}}
{\left( \frac{d-1}{2} \right)_{r+s}}~\frac{\left( \frac{1}{2} \right)_s}
{\left( 1 \right)_s} ~x^r y^s,
\nonumber\\
F_2\left(\frac{d-3}{2},1,1,\frac{3}{2},\frac{d-2}{2};x,y\right)&=&
\sum_{r=0}^{\infty} \sum_{s=0}^{\infty} 
\frac{\left( \frac{d-3}{2} \right)_{r+s}}
{\left( \frac{3}{2} \right)_r \left( \frac{d-2}{2} \right)_s} ~x^r y^s. 
\end{eqnarray}
The function $\phi(x,y)$ is
\begin{equation}
\phi(x,y)=
F^{1;2;1}_{1;1;0} \left[^{\frac{d-3}{2}:~ \frac{d-3}{2},~1;~~~~ 1;}
_{\frac{d-1}{2}:~~~~~~ \frac{d-2}{2};~~-;}~~x,y\right]=
\sum_{r=0}^{\infty} \sum_{s=0}^{\infty} 
\frac{\left( \frac{d-3}{2} \right)_{r+s}}
{\left( \frac{d-1}{2} \right)_{r+s}}~\frac{\left( \frac{d-3}{2} \right)_r}
{\left( \frac{d-2}{2} \right)_r} ~x^r y^s ,
\end{equation}
where $F^{1;2;1}_{1;1;0} $ is the Kamp\'e de F\'eriet function 
\cite{ApKdF}. The Appell function $F_1$ admits a one-fold integral
representation (see Appendix). The functions $F_2$ and $\phi$
admit two-fold integral representations, and this is the reason
why their $\varepsilon$ expansions are problematic.

We discovered that both functions can be represented in terms of 
the Gauss hypergeometric function $_2F_1$ and Appell function  $F_3$, 
defined as \cite{ApKdF}:
\begin{equation}
\label{eq:f3_def}
F_3\(\al,\ap,\bt,\bp,\ga,x,y \)
= \sum_{m,n=0}^{\infty} \frac{(\al)_m(\ap)_n(\bt)_m(\bp)_n
}{(\ga)_{m+n}} \, \frac{x^m}{m!}\, \frac{y^n}{n!},\qquad
|x|<1,\quad|y|<1.
\end{equation}

We found two methods to  obtain such a representation for the $\phi$ 
function. The first method is as follows. We write the function $\phi(x,y)$
as
\begin{equation}
\phi(x,y)=\sum_{r=0}^{\infty}
\frac{\left(\frac{d-3}{2}\right)_r 
  \left(\frac{d-3}{2}\right)_r}
{\left(\frac{d-2}{2}\right)_r
 \left(\frac{d-1}{2}\right)_r}
x^r \Fh21\Fey{1,\frac{d-3}{2}+r}{\frac{d-1}{2}+r},
\end{equation}
and then perform an analytic continuation of $_2F_1$ transforming it to
two functions $_2F_1$ with argument $1/y$. Thus, we  obtain 
two terms. One of these terms is just the
Gauss function $_2F_1$ and another one is the  Horn function
\begin{equation}
H_2\left(\frac{d-5}{2},1,1,1,\frac{d-2}{2},x,\frac{1}{y-1}\right),
\label{h2_for_phi}
\end{equation}
defined as \cite{Erdely}
\begin{equation}
\label{eq:h2_def}
H_2\(\al,\bt,\ga,\gp,\de,x,y\) 
= \sum_{m,n=0}^{\infty} \frac{(\al)_{m-n}(\bt)_m(\ga)_n(\gp)_n}
{(\de)_m}\, \frac{x^m}{m!} \, \frac{y^n}{n!},
\qquad\frac{1}{|y|}-|x|>1,~~|x|,|y|<1.
\end{equation}
The Horn function (\ref{h2_for_phi}) can be expressed in terms of 
$_2F_1$ functions  and the Appell function $F_3$ using Eq.~(65) 
on p.~295 in Ref.~\cite{Srivastava_Karlsson}. Combining all terms,
we arrive at the following result:
\begin{eqnarray}
\phi(x,y)&=&
-\frac{(d-3)(d-4)}{(d-5)(d-7)}\frac{1}{x(1-y)}
F_3\left(1,1,3-\frac{d}{2},1,\frac{9-d}{2};\frac{1}{x},
\frac{1}{1-y}\right)
\nonumber \\
&&{}
+\frac{(3-d)\Gamma\left(\frac{d-2}{2}\right)
\Gamma\left(\frac{5-d}{2}\right)}
{\sqrt{\pi}~(1-y)(-x)^{\frac{(d-5)}{2}}}
\sqrt{1-\frac{1}{x}}
~~\Fh21\FXY{1,1}{\frac32}
\nonumber \\
&&{}
+\frac{\Gamma\left(\frac{d-1}{2}\right)
\Gamma\left(\frac{5-d}{2}\right)}
{(-y)^{\frac{d-3}{2}}}
~\Fh21\Fxy{1,\frac{d-3}{2}}{\frac{d-2}{2}}.
\end{eqnarray}
The  Appell function $F_3$ for this specific set of parameters
can be written as a one-fold integral:
\begin{equation}
F_3 \left(1,1,3-\frac{d}{2},1,\frac{9-d}{2};\frac{1}{x},
\frac{1}{1-y} \right)=
\frac{\Gamma \left( \frac{9-d}{2} \right) x (y-1) }
{\Gamma \left( \frac32 \right) \Gamma \left(3-\frac{d}{2}\right)}
~ \int_0^1 \frac{(1-v)^{2-\frac{d}{2}}
\arcsin \sqrt{\frac{v}{1-y}}}
{\left( 1-x- v \right) \sqrt{1-y - v}}~dv.
\end{equation}
This formula can be obtained from the integral representation
of the  $F_3$ function \cite{Erdely} (see also Appendix).
The function $\phi(-m^2z,\theta)$ from Eq.~(\ref{solution})
reads:
%
%
\begin{eqnarray}
\phi(-m^2z,\theta) &=&(d-3) \Gamma\left(\frac{d-5}{2}\right)
\left\{ \frac{-\phi_1(s,t)}{\sqrt{\pi}\Gamma\left(2-\frac{d}{2}\right)}
\right.
\nonumber \\
&&{}+\frac12 \Gamma\left(\frac{d-3}{2}\right) (-\theta)^{\frac{3-d}{2}}
\Fh21\Fus{1,\frac{d-3}{2}}{\frac{d-2}{2}}
\nonumber \\
&&{}-\left.
\frac{t(u+t)\Gamma\left(\frac{d}{2}-1\right)}{2\sqrt{m^2stu}}
\left[\frac{t(t+u)}{4m^2u}\right]^{\frac{d-5}{2}}
\arcsin \sqrt{\frac{w}{4m^2(t+u)}}
\right\},
\end{eqnarray}
%
%
where
\begin{equation}
\phi_1(s,t)=\int_0^1 \frac{ (1-v)^{2-\frac{d}{2}}
\arcsin \sqrt{\frac{vt}{4m^2}}}
{\left[ \frac{w}{4(u+t)}- v \right] \sqrt{1-\theta - v}}~dv,
\end{equation}
and
\begin{equation}
w=16m^4-4m^2s-ts.
\label{omega}
\end{equation}
From this expression, it follows  that any coefficient in the 
$\varepsilon$ expansion of the $\phi$ function can be expressed in terms of 
one-fold integrals. Several terms of the $\varepsilon$ expansion 
of the function $\phi$ are given in Ref.~\cite{Fleischer:2003bg}.

We present also another method to represent the function $\phi(x,y)$ 
in terms of the Appell function $F_3$ and the Gauss functions $_2F_1$.
For the function $\phi(x,y)$, one can write the following 
integral representation \cite{Fleischer:2003bg}:
\begin{equation}
\phi(x,y)=\frac{d-3}{2}~\int \limits_0^1 
\frac{v^{\frac{d-5}{2}}}{1-yv}
\Fh21\FXV{1,\frac{d-3}{2}}{\frac{d-2}{2}}~dv~.
\label{phi_int_rep}
\end{equation}
From this integral representation, one can derive a differential
equation. Differentiating both sides of Eq.~(\ref{phi_int_rep}) 
w.r.t.\ $x$, using the following formula for the derivative of the 
Gauss function,
\begin{equation}
\frac{d}{dz} \Fh21\Fe{1,b}{c}=
\frac{(b z-c+1)}{z(1-z)}
\Fh21\Fe{1,b}{c}
 +\frac{(c-1)}{z(1-z)},
\end{equation}
and after some simplification of the resulting integrand, one
obtains the following equation \cite{Fleischer:2003rm}: 
\begin{eqnarray}
2(y-x)x\frac{\partial \phi(x,y)}{\partial x}
&=&[y -(d-3)(y-x)] \phi(x,y)
\label{equ_4_fi}
\\
&&{}-x(d-3)_2F_1\left(1,\frac{d-3}{2}, \frac{d-2}{2},x \right)
+(d-4)y~_2F_1\left(1,\frac{d-3}{2}, \frac{d-1}{2},y \right).
\nonumber
\end{eqnarray}
This first-order differential equation can be easily solved  
yielding:
\begin{eqnarray}
\phi(x,y)&=&\frac{(d-3)}{(d-2)}~\frac{x}{x-y}~
F_3\left(\frac12,1,1,\frac{d-3}{2},\frac{d}{2};
~\frac{x}{x-y},x\right)
\nonumber\\
&&{}+\left(\frac{y}{y-x}\right)^{1/2}
\Fh21\Fey{1,\frac{d-3}{2}}{\frac{d-1}{2}}
\Fh21\Fxy{1,\frac{d-3}{2}}{\frac{d-2}{2}}.
\label{fi_f3}
\end{eqnarray}
As it happens in the previous case,  instead of the Kamp\'e de F\'eriet
function, we obtain the  more familiar Appell function $F_3$, which,
for the above  parameters, admits the one-fold integral 
representation:
\begin{equation}
F_3\left(\frac12,1,1,\frac{d-3}{2},\frac{d}{2};
~\frac{x}{x-y},x\right)=
\frac{\Gamma\left(\frac{d}{2}\right)(x-y)}
{\sqrt{\pi}~\Gamma\left(\frac{d-3}{2}\right)x(1-x)}
\int_0^1 \frac{(1-v)^{\frac{d-5}{2}}}{1-\frac{vx}{x-1}}
\ln \frac{1+\sqrt{\frac{x v}{x-y}}}
        {1-\sqrt{\frac{x v}{x-y}}}~d v.
\label{secondF3}
\end{equation}
This relation is obtained from the integral representation
given in the Appendix. By using Eqs.~(\ref{fi_f3}) and 
(\ref{secondF3}), we obtain the  following expression
for the  function $\phi(-m^2z,\theta)$ 
from Eq.~(\ref{solution}):
%
%
\begin{eqnarray}
\phi(-m^2z,\theta)&=&
\frac{(d-3) \Gamma\left(\frac{d}{2}-1\right)
(4m^2-s) t }
{2 \sqrt{\pi}~\Gamma\left(\frac{d-3}{2} \right)
w }
\phi_2(s,t)
\nonumber \\
&&{}+\left[\left(1-\frac{4m^2}{s}\right)
         \theta \right]^{\frac12}~
\Fh21\Ftt{1,\frac{d-3}{2}}{\frac{d-1}{2}}
\Fh21\FTT{1,\frac{d-3}{2}}{\frac{d-2}{2}},
\end{eqnarray}
where 
\begin{equation}
\phi_2(s,t)=\int_0^1\frac{(1-v)^{\frac{d-5}{2}}}
{1-\frac{4m^2uv}{w}}
\ln\frac{1+\sqrt{-\frac{4m^2uv}{ts}}}
        {1-\sqrt{-\frac{4m^2uv}{ts}}} dv,
\end{equation}
%
%
$\theta$ is defined in Eq.~(\ref{z_teta}) and $w$ in Eq.~(\ref{omega}).

Thus, we obtained two rather different hypergeometric  representations
for the $\phi$ function. In both cases, the hypergeometric functions
admit one-fold integral representations, so that all the coefficients in the
$\varepsilon$ expansion may
be expressed  only in terms of one-fold integrals.

The Appell function $F_2$ from Eq.~(\ref{solution}) can also
be expressed in terms of the Appell function $F_3$ and the Gauss
function $_2F_1$.
To obtain such a relation, we use the formula for the analytic
continuation of the Appell function $F_3$ from Ref.~\cite{Erdely} and
obtain:
\begin{eqnarray}
\lefteqn{F_2\left(\frac{d-3}{2},1,1,\frac{d-2}{2},\frac32;x,y\right)=
\frac{(d-4)}{(d-5)(d-7) x y}~
F_3\left(1,1,3-\frac{d}{2},\frac12,\frac{9-d}{2};
\frac{1}{x},\frac{1}{y}\right)}
\nonumber \\
&&{}
-\frac{\sqrt{\pi} ~ \Gamma\left(\frac{5-d}{2}\right)
\Gamma\left(\frac{d-2}{2}\right) }
{2\sqrt{-y} (-x)^{\frac{d}{2}-2} }
~+~\frac{ \Gamma\left(\frac{5-d}{2}\right)
\Gamma\left(\frac{d-2}{2}\right) }{\sqrt{\pi}~
(-x)^{\frac{d}{2}-2}  \sqrt{1-x} }
~\Fh21\FxC{1, \frac12}{\frac32 }
\nonumber
\\
&&{}
 + \frac{\sqrt{\pi} ~\Gamma\left(\frac{5-d}{2}\right) }
{2\Gamma\left(\frac{6-d}{2}\right) \sqrt{-y} ~(1-y)^{\frac{d}{2}-2} }
~\Fh21\FxP{1,\frac{d}{2}-2}{\frac{d}{2}-1}.
\end{eqnarray}
The Appell function $F_3$ with this particular set of parameters
also can be expressed in terms of the one-fold integral
\begin{equation}
F_3\left(1,1,3-\frac{d}{2},\frac12,\frac{9-d}{2};
\frac{1}{x},\frac{1}{y}\right)=
\frac{-\Gamma\left(\frac{9-d}{2}\right)}
{\sqrt{\pi}~\Gamma\left(3-\frac{d}{2}\right)}
~\frac{x \sqrt{y}}{1-x}
\int_0^1\frac{(1-v)^{\frac{4-d}{2}}}
{1+\frac{v}{x-1}}
\ln \frac{1+\sqrt{\frac{v}{y}}}{1-\sqrt{\frac{v}{y}}}~dv.
\end{equation}
Therefore, the Appell function $F_2$ from Eq.~(\ref{solution}) reads:
%
%

\begin{eqnarray}
\lefteqn{F_2\left(\frac{d-3}{2},1,1,\frac32, \frac{d-2}{2};
 \frac{s}{s-4m^2}, -m^2z \right)}
\nonumber \\
&=&
\frac{ \Gamma\left(\frac{5-d}{2}\right)}
   {16 \sqrt{\pi}  \Gamma\left(2-\frac{d}{2}\right)}
  \frac{(4 m^2-s)}{m^2} 
  \left[{t(s-4m^2)}{m^2 u}\right]^{\frac12}
  \phi_3(s,t)
\nonumber \\
&&{}
+ \Gamma\left(\frac{5-d}{2} \right)
 \Gamma\left(\frac{d}{2}-1 \right)
 \left[\frac{\pi s t}{16 u m^2}\right]^{\frac12}
 \left[\frac{4m^2-s}{s}\right]^{\frac{d-3}{2}}
\left[ 1+ \frac{i}{\pi}~
\ln \frac{1+\sqrt{- \frac{u}{t}}}{1-\sqrt{- \frac{u}{t}}}
\right]
\nonumber \\
&&{}+  \frac{\sqrt{\pi}~ \Gamma\left(\frac{5-d}{2}\right)}
         {4 \Gamma\left(3-\frac{d}{2}\right)}
	\left[ \frac{t ( 4 m^2-s)}{u m^2}\right]^{\frac12}  
        \left[\frac{t(4m^2-s)}{w}\right]^{\frac{d-4}{2}}
        \Fh21\FmW{1,\frac{d-4}{2}}{\frac{d-2}{2}},
\end{eqnarray} 
where
\begin{equation}
\phi_3(s,t)=\int_0^1 \frac{(1-v)^{\frac{4-d}{2}}}
           {1+\frac{(s-4m^2)v}{4m^2}}
  \ln \frac{1+\sqrt{\frac{vt(s-4m^2)}{4m^2u} }}
           {1-\sqrt{\frac{vt(s-4m^2)}{4m^2u} }}~
	   d v.
\end{equation}

Thus, we found that the Appell function $F_2$, the function
$\phi$  and, therefore, also the integral $B(s,t)$ are 
expressible in terms of hypergeometric functions
admitting one-fold integral representations.
Such a representation of $B(s,t)$  would be convenient for obtaining
higher-order terms in the $\varepsilon$ expansion of this
integral. The first terms in the $\varepsilon$ expansion
of the integral $B(s,t)$ were obtained in 
Ref.~\cite{Bonciani:2003cj}.

In Ref.~\cite{Fleischer:2003rm}, a  one-fold integral 
representation for a scalar box integral with arbitrary
masses, external momenta and  space-time dimension
$d$ was presented. To obtain the formula for $B(s,t)$ directly from 
such a  representation by setting masses and scalar invariants to
their specific values would be not so easy because the appropriate analytic 
formulae are rather lengthy, and also the analytic continuations needed for 
the hypergeometric functions are rather nontrivial. Our result can be 
considered as a confirmation that the one-fold integral representations 
for box integrals with physically relevant  kinematics do exist.

\section{Imaginary parts and spectral representation for the 
integral \boldmath{$D_2$}}

The integral $I_4^{(d)}$ with all internal lines massless and
external legs on shell can be calculated analytically.
Using the method of dimensional recurrences 
\cite{Tarasov:2000sf}, we obtain the following relation
for the last integral in Eq.~(\ref{alpha_minus}),
assuming  $|s+t| \leq |s|$ and $|s+t| \leq |t|$:
\begin{eqnarray}
\lefteqn{I_4^{(d)}(0,0,0,0;~0,0,0,0;~s,t)=\frac{-4(d-3)}{s t(d-4)}}
\nonumber
\\
&&{}
\times
     \left\{ I_2^{(d)}(0,0;~t)
     \Fh21\Fis{1,\frac{d}{2}-2}{\frac{d}{2}-1}
           +I_2^{(d)}(0,0;~s)
     \Fh21\Fit{1,\frac{d}{2}-2}{\frac{d}{2}-1}
         \right\}.
\label{spt_leq_s_t}
\end{eqnarray}
This formula is in agreement with the result obtained in 
Ref.~\cite{Bern:1993kr} (see also
Refs.~\cite{Duplancic:2000sk,Kurihara:2005at}).
Thus, Eqs.~(\ref{D2_B_I4}), (\ref{solution}), and (\ref{spt_leq_s_t}) 
provide us with a hypergeometric representation for the integral
$D_2$.
The  $\varepsilon$ expansions of the real and imaginary parts  
of this integral through order  $\varepsilon^2$ were given in 
Ref.~\cite{Korner:2004rr}.
We expect that, with our hypergeometric representation,
one can derive a shorter result for the $\varepsilon^2$ term
in the expansion of the integral $D_2$ than that given
in Ref.~\cite{Korner:2004rr}.


The functional equations in Eqs.~(\ref{analitic_cont}) and (\ref{alpha_minus})
can be used for finding imaginary parts of the integrals
in some kinematical regions.
As one can see from Eq.~(\ref{analitic_cont}),  if  $s>m^2$ and $t<0$, 
the integral  $I_4^{(d)}(0,0,0,m^2;~0,0,m^2,m^2;~s,t)$ has
an imaginary part that arises only from the integral
$I_4^{(d)}$  with all propagators massless, 
which can be  easily found from Eq.~(\ref{spt_leq_s_t}), so that
\begin{eqnarray}
 {\rm Im }~D_2(t,s)
=\frac{4(d-3)}{(d-4)} \frac{\sin{\frac{\pi d}{2}}}
{t(m^2-s)}
I_2^{(d)}\left(0,0;~\frac{(s-m^2)^2}{-s}\right)
\Fh21\FHQ{1,\frac{d}{2}-2}{\frac{d}{2}-1}, &&
\nonumber
\\
s>m^2,\quad t<0.&&
\label{impart_d2_s}
\end{eqnarray}
In a similar fashion, if $s>4m^2$ and $t<0$, the imaginary part
of the integral $B(s,t)$  can be  obtained from 
Eq.~(\ref{alpha_minus}).  In this case, the imaginary part 
originates only from the second integral on the right-hand side of 
Eq.~(\ref{alpha_minus}). Again one can use Eq.~(\ref{spt_leq_s_t})
to find:
\begin{eqnarray}
{\rm Im }~ B(s,t)
=\frac{4(d-3)}{(d-4)} \frac{\beta_s~
\sin\frac{\pi d}{2} }
{t(4m^2-s)} I_2^{(d)}\left(0,0;~4m^2-s \right)
\Fh21\FBB{1,\frac{d}{2}-2}{\frac{d}{2}-1},&&
\nonumber \\
s>4 m^2,\quad t<0.&&
\end{eqnarray}
One can write  $D_2(s,t)$ in the fixed-$t$ spectral representation
\begin{equation}
D_2(t,s)= \frac{1}{\pi} \int \limits_{m^2}^{\infty}
\frac{dx\,{\rm Im}~D_2(t,x)}{s-x}.
\label{d2_fixed_t}
\end{equation}
Substituting Eq.~(\ref{impart_d2_s}) into Eq.~(\ref{d2_fixed_t}) 
leads to the following  integral representation for $D_2(t,s)$:
\begin{eqnarray}
D_2(t,s)&=&\frac{4(d-3) \sin { \frac{\pi d}{2}}}
{(d-4)t \pi} \int \limits_{m^2}^{\infty}\frac{dx}{(s-x)(x-m^2)}
\nonumber \\
&&{}\times
~I_2^{(d)}\left(0,0;~\frac{(x-m^2)^2}{-x}\right)
\Fh21\FSX{1,\frac{d}{2}-2}{\frac{d}{2}-1}.
\label{int_rep_d2}
\end{eqnarray}
To use $D_2(t,s)$ in calculations of heavy-quark production,
one needs to know it at $t>0$ and $s<0$. 
To perform  the analytic  continuation of $D_2(t,s)$ into the region 
$t>0$, one can use  Eq.~(\ref{int_rep_d2}). 
Using a formula for the analytic continuation of the Gauss
hypergeometric function \cite{Erdely} and introducing the new integration variable
\begin{equation}
v =\frac{xt}{(m^2-x)^2 + xt},
\end{equation}
leads to the following expression:
\begin{eqnarray}
D_2(t,s)&=&\frac{2(d-3)  \sin { \frac{\pi d}{2}}}
{\pi s t^{4-\frac{d}{2}}} \int \limits_0^1\frac{~d v}
{1-v \sigma  }
~I_2^{(d)}\left(0,0;~\frac{v-1}{v}\right)
 \left\{\frac{m^2-s}{1-v}
 -\frac{(s+m^2)}{\sqrt{(1-v)(1-v \theta )}}
\right\}
\nonumber \\
&&{}\times
\left\{\frac{v}{6-d}~
\Fh21\FSW{1,3-\frac{d}{2}}{4-\frac{d}{2}}
+\frac{\pi}{2}  v^{\frac{d}{2}-2}
\cot { \frac{\pi d}{2}}     
+i\frac{\pi}{2}v^{\frac{d}{2}-2} \right\}, 
\end{eqnarray}
where
\begin{equation}
\sigma= 1+ \frac{(s-m^2)^2}{st},
\end{equation}
and $\theta$ is defined in  Eq.~(\ref{z_teta}).
Substituting the expression for $I_2^{(d)}$ in this formula, 
we obtain for the imaginary part:
\begin{eqnarray}
{\rm Im}~D_2(t,s)=\frac{-\pi^{\frac32}(d-3)t^{\frac{d}{2}-4}}
{s 2^{d-4}\Gamma\left(\frac{d-1}{2}\right)}
\left\{\frac{m^2-s}{d-4}~\Fh21\Fsg{1,1}{\frac{d-2}{2}}
-\frac{s+m^2}{d-3}F_1\left(1,1,\frac{1}{2};\frac{d-1}{2};
\sigma, \theta \right)
\right\},&&
\nonumber\\
 s<0,\quad t>0.&&
\nonumber\\
\label{imag_part_d2}
\end{eqnarray}
Applying the Euler transformation to the hypergeometric functions
$_2F_1$ and $F_1$ (see, for example, Ref.~\cite{Erdely}), we find 
agreement with the expression for the imaginary part obtained in 
this region from the analytic result for the integral $B(s,t)$ given
in Eq.~(\ref{solution}).

For the real part, one can write the following 
representation:
\begin{eqnarray} 
{\rm Re}~D_2(t,s)=
\frac{2(d-3)(1-\sigma)^{\frac{d}{2}-2}}
{(d-4)~t(s-m^2)} I_2^{(d)}(0,0;-t)
\Fh21\Fsg{1,\frac{d-4}{2}}{\frac{d-2}{2}}
+ \cot \frac{\pi d}{2} ~~{\rm Im}~D_2(t,s)
+ \Phi(t,s)
,&&
\nonumber \\
s<0,\quad t>0,&&
\nonumber\\
\label{real_part_d2}
\end{eqnarray}
where
\begin{equation}
\Phi(t,s)=\frac{2^{4-d}~ \sqrt{\pi}~ (d-3) (s+m^2) }
{(6-d) \Gamma\left(\frac{d-1}{2} \right) s ~t^{4-\frac{d}{2}}}
 \int \limits_0^1\frac{d v}
{1-v \sigma  }
~ \frac{  v^{3-\frac{d}{2}}(1-v)^{\frac{d-5}{2}} }
 {\sqrt{(1-v \theta)}}
\Fh21\FSW{1,3-\frac{d}{2}}{4-\frac{d}{2}}.
\label{big_fi}
\end{equation}
Equations~(\ref{imag_part_d2}) and (\ref{real_part_d2}) are convenient
for the  $\varepsilon$  expansion.  Appropriate formulae for
the $\varepsilon$  expansion of the Gauss  hypergeometric function 
can be taken from Refs.~\cite{Davydychev:2000na, Kalmykov:2008ge}.
To obtain the  $\varepsilon$ expansion of the Appell hypergeometric 
function $F_1$, one can use a one-fold  integral representation
(see Appendix).
The first two terms in Eq.~(\ref{real_part_d2})  give singular
contributions proportional to $1/\varepsilon^2$ and $1/\varepsilon$.
The last term in Eq.~(\ref{real_part_d2}) is regular in $\varepsilon$.
Its $\varepsilon$ expansion can be derived from integral representation
of Eq.~(\ref{big_fi}).

In the case when $t>0$ and $s<0$, the imaginary part of $B(s,t)$ can be found
from Eq.~(\ref{solution}) and reads:
\begin{equation}
{\rm Im}~B(s,t)= \frac{ \pi^{\frac32}~ 2^{4-d}~t^{\frac{d-5}{2}}}
{m (s-4m^2) \Gamma\left( \frac{d-1}{2}\right)}
F_1\left(\frac{d-3}{2},1,\frac12; \frac{d-1}{2};
1-\frac{t}{4m^2-s},1-\frac{t}{4m^2}
\right).
\end{equation}

The imaginary part of the integral $D_2(s,t)$ for $s>0$ and $t<0$
reads:
\begin{eqnarray}
{\rm Im}~D_2(s,t)&=&\sin \frac{\pi d}{2}~I_2^{(d)}(0,0;~-s)~
\left\{
\frac{2(d-3)~ }{(d-4)s(m^2-t)}
~\Fh21\FHH{1,\frac{d}{2}-2}{\frac{d}{2}-1}
\right.
\nonumber \\
&&{}-\left.
\frac{(t+m^2) }
{m \sqrt{s} (t-m^2)^2}~
F_1\left(\frac{d-3}{2},1,\frac12, \frac{d-1}{2};
1+\frac{st}{(t-m^2)^2} 
,1-\frac{s}{4m^2}
\right)\right\}.
\end{eqnarray}
The first few terms of the $\varepsilon$ expansion
of the Appell function $F_1$ can be found in
Refs.~\cite{Fleischer:2003rm,Davydychev:2005nf}.

\section{Functional equations for \boldmath{$D_2$} with one quark 
leg off shell}

For the computation of the radiative corrections to  heavy-quark 
production in the NNLO approximation, one needs to know
the integral $D_2$ with one quark leg off shell. This is a 
rather nontrivial task,  which will be considered in a
forthcoming publication. In this section,  we just want to 
outline the  strategy  of such calculations. Similar to the case 
of $D_2$  with all legs on shell, one can derive the functional 
equations:
\begin{eqnarray}
\lefteqn{I_4^{(d)}(0,0,0,m^2;0,0,s_{34},m^2;~s,t)}
\nonumber \\
&=& 
\frac{m^2}{s} 
I_4^{(d)}\left(0,0,0,m^2;0,\frac{(m^2-s_{34})(m^2-s)}{s},
 s_{34},m^2,\frac{m^4}{s},t\right)
\nonumber \\
&&{}+
\frac{s-m^2}{s} I_4^{(d)}\left(0,0,0,0;
~\frac{(m^2-s_{34})(m^2-s)}{s},
 0,0,0;~ \frac{(m^2-s)^2}{s},t\right),
\label{inverse_s}\\
\lefteqn{I_4^{(d)}(0,0,0,m^2;~0,0,s_{34},m^2;~s,t)}
\nonumber\\
&=&
\frac{m^2}{s_{34}}
I_4^{(d)}\left(0,0,0,m^2;~\frac{(m^2-s_{34})(m^2-s)}{s_{34}},0,m^2,
  \frac{m^4}{s_{34}};~s,\frac{t m^2}{s_{34}}\right)
\nonumber\\
&&{}+\frac{s_{34}-m^2}{s_{34}}
I_4^{(d)}\left(0,0,0,0;~\frac{tm^2}{s_{34}},
\frac{(m^2-s_{34})^2}{s_{34}},0,0;
~\frac{(m^2-s_{34})(m^2-s)}{s_{34}},t\right).
\label{inverse_s34}
\end{eqnarray}
Equations (\ref{inverse_s}) and (\ref{inverse_s34}) were obtained from
Eq.~(\ref{I5intoI4}) by setting 
\begin{eqnarray}
m_1^2&=&m_2^2=m_3^2=m_5^2=s_{12}=s_{23}=0,\qquad m_4^2=s_{14}=m^2,
\nonumber \\
G_4&=&\partial_5\Delta_5=\partial_1\partial_5 \Delta_5=0,
\end{eqnarray}
in both cases and additionally 
\begin{equation}
\partial_3\partial_5 \Delta_5=0,
\end{equation}
in Eq.~(\ref{inverse_s})
and 
\begin{equation}
\partial_2\partial_5 \Delta_5=0,
\end{equation}
in Eq.~(\ref{inverse_s34}). The above equations are rather similar
to the functional equations for the integral $D_2$ with all legs 
on shell. In some kinematical regions, the  imaginary part of the 
integral $D_2(s_{34},s,t)$ arises from integrals with all propagators
massless. In Eq.~(\ref{inverse_s}), the integral with massless 
propagators can be expressed in terms of three Gauss  hypergeometric
functions \cite{Bern:1993kr}. The analytic expression for the massless 
integral in Eq.~(\ref{inverse_s34}) for arbitrary $d$ is not known 
at present. The first terms in the $\varepsilon$ expansion of both 
integrals may be  found in Refs.~\cite{Bern:1993kr,Duplancic:2000sk}.
We expect that, similar to the $D_2$ integral with all legs
on shell,  the integral $D_2$ with one leg off shell can be represented
in terms of hypergeometric functions admitting  a one-fold integral 
representation.

\section{Conclusions}

The usefulness of  functional equations turns out to be  threefold.
First, we obtain a hypergeometric representation of the integral 
needed for NLO calculations of heavy-quark production from the 
result for the integral from Bhabha scattering.
Second, since the $\varepsilon^2$ term in the expansion of the 
integral $D_2$  is known \cite{Korner:2004rr}, one can use it 
to obtain the $\varepsilon^2$ term for Bhabha scattering.   
Third, functional equations provide a simple method to obtain
imaginary parts of integrals. For some kinematic regions, the 
imaginary parts of  integrals with  nonzero internal masses 
can be  related  to the integrals with all lines massless.  

It is also important  that the functional equations provide a  
tool for performing analytic continuations of the considered 
integrals. As was already observed in Ref.~\cite{Tarasov:2008hw}  
and now also in this paper, functional equations suitable for analytic 
continuation express the considered integral in terms of the same 
integral with transformed arguments that has no imaginary part  
plus simpler  integrals (usually with massless  lines) giving the 
imaginary part of the integral. Such analytic continuation is achieved 
by  solving algebraic equations, so that the explicit analytic form 
of the integral  is not needed.


\section{Acknowledgments}
We are grateful to Z.~Merebashvili for useful discussions.
This work was supported in part by BMBF Grant No.~HT6QUA
and DFG Grant No.~KN365/3-2.

\section{Appendix}

In this appendix, we present formulae which were used
in the derivation of some equations in the main text.

The integral representation of the Appell hypergeometric
function $F_1$ reads:
\begin{eqnarray}
F_1(\alpha,\beta,\beta',\gamma;~x,y)=
\frac{\Gamma(\gamma)}{\Gamma(\alpha)\Gamma(\gamma-\alpha)}
\int \limits_0^1\frac{~u^{\alpha-1}(1-u)^{\gamma-\alpha-1}}
{(1-ux)^{\beta}(1-uy)^{\beta'}}du,&&
\nonumber\\
{\rm Re}~\alpha>0,\quad {\rm Re}~(\gamma-\alpha)>0.&&
\end{eqnarray}
The
integral representation of the Appell function $F_3$ used
in the  derivation of the one-fold integral representations of
the $\phi_i$ functions reads:
\begin{equation}
F_3(\alpha,\alpha',\beta,\beta',\gamma; x, y)=
\frac{\Gamma(\gamma)}{\Gamma(\gamma-\beta)
\Gamma(\beta)}
\int_0^1\frac{u^{\gamma-\beta-1}(1-u)^{\beta-1}}
{(1-x+ux)^{\alpha}}
\Fh21\Fuy{\alpha',\beta'}{\gamma-\beta}~du.
\end{equation}
This integral representation follows from Eq.~(20)
in Ref.~\cite{Prudnikov:1986}.

\end{document}